\documentclass[prd,aps,nofootinbib,floatfix,10 pt]{revtex4}
\usepackage[T1]{fontenc} 
\usepackage{amsmath,graphicx,color,epsfig}
\usepackage{subfig}
\usepackage{epsfig}
\usepackage{pstricks}
\usepackage{float}
\usepackage{tensor}
\usepackage{gensymb}

\begin{document}


\title{Neutrino Mass Matrix in a gauge group $SU(2)_L \times U(1)_e \times U(1)_\mu \times U(1)_\tau$}
\author{Fayyazuddin}
\affiliation{National Centre for Physics, Quaid-i-Azam University Campus, Islamabad 45320, Pakistan.}

\date{\today}

\begin{abstract}
The electroweak unification group $G\equiv SU(2)_L\times U(1)_e\times U(1)_\mu\times U(1)_\tau$ in which each fermion multiplet has its own $U(1)$ factor was proposed in 1986 to get the neutrino mass matrix. In this paper, the gauge group G is restricted to lepton section only, leaving quark multiplets as in the standard model. In addition to lepton multiplets $L_e$, $L_\mu$ and $L_\tau$, there are three $SU(2)$singlet right handed neutrinos $N_{R}^{(i)}$'s. WIth the breaking of G to $SU(2)_L\times U(1)$, the right handed neutrinos acquire heavy Majorana masses. Three heavy right handed neutrinos $N_{R}^{(i)}$'s are available to generate a $3\times 3$ non-diagonal neutrino mass matrix in terms of three Yukawa couplings $h^{(2)}_{1}$, $h^{(3)}_{2}$, $h^{(1)}_{3}$ of the Higgs scalar doublet to $L_e$, $L_\mu$, $L_\tau$ with $N_{R}^{(1)}$, $N_{R}^{(2)}$ and $N_{R}^{(3)}$ respectively. Three Yukawa couplings can be arranged and expressed in terms of masses $m_e$, $m_\mu$, $m_\tau$ in three different ways to obtain the results of interest for Case 1: ($\nu_e \rightarrow \nu_\tau$); Case 2: ($\nu_e \rightarrow \nu_\mu$); Case 3: ($\nu_\mu \rightarrow \nu_\tau$). The results obtained for the three cases are compared with the experimental data from neutrino oscillations. Cases 1 and 2 are relevant for solar neutrino oscillations whereas Case 3 is relevant for atmospheric neutrino oscillations.
\end{abstract}

\maketitle

\section{ INTRODUCTION}

In 1986 \cite{1}, a simple extension of the standard electroweak gauge group to the group $G\equiv SU(2)_L\times U(1)_e\times U(1)_\mu\times U(1)_\tau$ in which each fermion multiplet of the standard model has its own U(1) factor was purposed to get the neutrinos masses.

The neutrino oscillations require that neutrinos are not massless but have very small masses \cite{2,3}. In 1986 - 1987, no accurate data of neutrino masses was available. It is appropriate to revisit the model to confront its predictions with the recent data about neutrino masses.

Since main predictions of gauge group G pertain to neutrino masses, the extension of the standard model gauge group to the gauge group G is restricted to only lepton section, leaving the quark multiplets as in the standard model.

Thus in the gauge group $G$, the lepton multiplets are assigned as follows \cite{1}

\begin{eqnarray}
L^i \equiv L_e =    \left(\begin{array}{c} \nu_e \notag \\ 
e \end{array}\right)_L : \left(\begin{array}{cccc} 2, & -1, & 0, & 0 \end{array}\right)\notag \\ 
e_R = \hspace{4.5em}   : \left(\begin{array}{cccc} 1, & -2, & 0, & 0 \end{array}\right) \notag\\ 
L_\mu =    \left(\begin{array}{c} \nu_\mu \\ \mu \end{array}\right)_L : \left(\begin{array}{cccc} 2, & 0, & -1, & 0 \end{array}\right) \notag\\ 
\mu_R = \hspace{4.5em}   : \left(\begin{array}{cccc} 1, & 0, & -2, & 0 \end{array}\right)\notag\\ 
L_\tau =    \left(\begin{array}{c} \nu_\tau \notag\\ 
\tau \end{array}\right)_L : \left(\begin{array}{cccc} 2, & 0, & 0, & -1 \end{array}\right) \notag\\ 
\tau_R = \hspace{4.5em}   : \left(\begin{array}{cccc} 1, & 0, & 0, & -2 \end{array}\right)
\end{eqnarray}

In addition three right-handed neutrinos $N_{R}^{(i)}$'s which are $SU(2)$ singlet but carry $U(1)_i$ quantum numbers:
$$N^{(1)}_R:\left(\begin{array}{cccc} 1, & -1, & 1, & 0 \end{array}\right), N^{(2)}_R:\left(\begin{array}{cccc} 1, & 1, & 0, & -1 \end{array}\right), N^{(3)}_R:\left(\begin{array}{cccc} 1, & 0, & -1, & 1 \end{array}\right) $$
are introduced.

The gauge vector bosons of the group $G$:
\begin{eqnarray}
SU(2)_L \times U(1)_e \times U(1)_\mu \times U(1)_\tau \\  
g \hspace{3.5em} g_{1e}  \hspace{2.5em} g_{1\mu}  \hspace{2em} g_{1\tau}
\end{eqnarray}
are 
\begin{eqnarray}
W^{\pm,0}_\mu, B^{(1)}_\mu, B^{(2)}_\mu, B^{(3)}_\mu:~ B^{(i)}_\mu \hspace{3em} i = 1,2,3 
\end{eqnarray}

Using the $e \leftrightarrow \mu \leftrightarrow \tau$ discrete symmetry
\begin{eqnarray}
g_{1e}  ~=~ g_{1\mu}  ~=~ g_{1\tau} ~= ~ g_1  \\
Q = T_{3L} +\frac{1}{2} (Y_1 + Y_2 + Y_3) \\ 
\frac{1}{e^2} =   \frac{1}{g^2} + \frac{1}{g^2_{1e}} + \frac{1}{g^2_{1\mu}} + \frac{1}{g^2_{1\tau}}  \\
\frac{1}{e^2}  =   \frac{1}{g^2} +  \frac{3}{g^2_1} =  \frac{1}{g^2} + \frac{1}{g'^{2}}  
\\ g'^{2} = g^{2}_1 / 3
\end{eqnarray}
in the symmetry limit.

First we note that following combinations of $B^{(i)}_\mu$'s give the three physical gauge vector bosons of group G: 
\begin{eqnarray}
B_\mu = \frac{1}{\sqrt{3}} (B^{(1)}_\mu + B^{(2)}_\mu + B^{(3)}_\mu) \\
X_\mu = \frac{1}{\sqrt{2}} (B^{(1)}_\mu -  B^{(3)}_\mu) \\ 
X^{\prime}_\mu = \frac{1}{\sqrt{6}} [(B^{(1)}_\mu - B^{(2)}_\mu) - (B^{(2)}_\mu - B^{(3)}_\mu)] 
\end{eqnarray}

\section{INTERACTION LAGRANGIAN FOR LEPTONS}
The interaction lagrangian for leptons is given by \cite{1}

\begin{eqnarray}
\mathcal{L}_{int} &=& [\frac {-g}{2\sqrt{2}} (J^{\dagger \lambda} W^{\dagger}_{\lambda} + h.c.) - \frac{g_1}{\sqrt{2}} \sum_{i} J^{(i)\lambda}_3 W_{3\lambda} - \frac{g^\prime}{2} \sum_{i} J^{(i)\lambda}_1 B_\lambda] - \frac{g_1}{2\sqrt{2}} [ (J^{(e)\lambda}_1 - J^{(\tau)\lambda}_1) X_\lambda \notag  \\ 
&+& \frac {1}{\sqrt{3}}( J^{(e)\lambda}_1 - J^{(\mu)\lambda}_1 + J^{(\tau)\lambda}_1) X^{\prime}_\lambda]  
+ \frac{g_1}{2\sqrt{2}}[(\bar{N}^{(1)} \gamma^\lambda N^{(1)} - 2\bar{N}^{(2)} \gamma^\lambda N^{(2)} + \bar{N}^{(3)} \gamma^\lambda N^{(3)})_R X_\lambda \notag\\
&+& \sqrt{3}(\bar{N}^{(1)} \gamma^\lambda N^{(1)} -  \bar{N}^{(3)} \gamma^\lambda N^{(3)})_R X^\prime_\lambda] \label{Lint}
\end{eqnarray}
where
\begin{eqnarray}
J^{\dagger \lambda} = 2 [\bar{\nu}^{(i)}_L \gamma^{\lambda} e^{(i)}_L],~
J_{3}^{(i) \lambda} = 2 [\bar{\nu}^{(i)}_L \gamma^{\lambda} \nu^{(i)}_L - \bar{e}^{(i)}_L \gamma^{\lambda} e^{(i)}_L],~
J^{(i) \lambda}_1 = 2 [- (\bar{\nu}^{(i)}_L \gamma^{\lambda} \nu^{(i)}_L - \bar{e}^{(i)}_L \gamma^{\lambda} e^{(i)}_L) - 2 \bar{e}^{(i)} \gamma^{\lambda} e^{(i)}]
\end{eqnarray}

The first term in Eq(\ref{Lint}) is $\mathcal{L}_{int}$ for leptons as in the standard model.

In order to spontaneously break the group G to the standard model group, three Higgs singlet $\Sigma^{(i)}$'s are introduced.

$$\Sigma^{(1)} : \left(\begin{array}{cccc} 1, & 1, & -1, & 0 \end{array}\right), \Sigma^{(2)} : \left(\begin{array}{cccc} 1, & 1, & 0, & -1 \end{array}\right), \Sigma^{(3)} : \left(\begin{array}{cccc} 1, & 0, & 1, & -1 \end{array}\right)$$

The gauge group G is broken to $SU(2)_L \times U(1)_Y$ by giving the vacuum expectation values to Higgs bosons $\Sigma^{(i)} = \frac{V_i}{\sqrt{2}}$. With the breaking of the gauge group G, the vector bosons $X_\mu$ and $X^{\prime}_\mu$ acquire heavy masses leaving $B_\mu$ massless. Similarly right handed neutrinos $N^{(i)}_{R}$'s acquire heavy Majorana masses.

\begin{eqnarray}
\mathcal{L}_{mass} (X) &=& \frac{1}{8} g^2_1 \frac{1}{2} [ (V^2_1 + V^2_2 + V^2_3) X^\mu X_\mu+(V^2_1 + V^2_2) X^{\prime \mu} X^\prime_\mu   \notag\\ 
&+&  \sqrt{3} (V^2_1 - V^2_2) (X^\mu X^{\prime}_\mu + X^{\prime}_\mu X^\mu) ]
\end{eqnarray}

To eliminate the cross term, we take $V_1 = V_2 = V$ and put $V_3 = V^\prime$ 

\begin{eqnarray}
\mathcal{L}_{mass} (X) = \frac{1}{8} \frac{1}{2} \left[2(V^2 + V^{\prime 2}) X^\mu X_\mu + 6 V^2  X^{\prime}\mu X^{\prime}_\mu \right]
\end{eqnarray}

For simplicity, we take $V = V^\prime$, so that

\begin{eqnarray}
m^2_X = m^2_{X^{\prime}} = \frac{3}{4} g^2_1 V^2 
\end{eqnarray}

\begin{eqnarray}
\mathcal{L}_{mass} (N) = [ f_{12} (N^{T(1)} C^{-1} N^{(2)} + N^{T(2)} C^{-1} N^{(1)})_R \overline{\Sigma}^{(3)}  \notag\\
+ f_{13} (N^{T(1)} C^{-1} N^{(3)} + N^{T(3)} C^{-1} N^{(1)})_R \overline{\Sigma}^{(2)}  \notag\\
+ f_{23} (N^{T(2)} C^{-1} N^{(3)} + N^{T(3)} C^{-1} N^{(2)})_R \overline{\Sigma}^{(1)} + h.c. ]\label{Lmass}
\end{eqnarray}


From Eq(\ref{Lmass}), we get the Majorana mass matrix for $N^{(i)}_{R}$'s
\begin{eqnarray}
M_R =
  \left[ {\begin{array}{ccc}
   0 & f_{12} & f_{13} \\
   f_{12} & 0 & f_{23} \\
   f_{13} & f_{23} & 0 \\   
   \end{array} } \right] \label{M_R}
\end{eqnarray}

To conclude, we note that after, spontaneously breaking the gauge group G to $SU(2)_L \times U(1)_Y$, the heavy vector bosons $X_\mu, X^{\prime}_\mu$ and heavy right handed neutrinos $N^{(i)}_{R}$'s are decoupled. The lepton multiplets along with the quark multiplets give the complete fermion contents of the standard model $SU(2)_L \times U(1)_Y$ with Y = -1 and Y = $\frac{1}{3}$ for the lepton and quark doublet of $SU(2)_L$ respectively. The first term in Eq(\ref{Lint}) supplemented with $\mathcal{L}_{int}$(quark) gives $\mathcal{L}_{int}$ of the standard model with gauge vector bosons $W^{\pm}_\mu, Z_\mu, A_\mu$

\section{NEUTRINO MASS MATRIX}
For simplicity, we take $f_{12} = f_{13} = f_{23} = f $ in Eq(\ref{M_R}), to give the mass matrix \cite{1}

\begin{eqnarray}
   M_R = \frac{f_V}{\sqrt{2}}
  \left[ {\begin{array}{ccc}
   0 & 1 & 1 \\
   1 & 0 & 1 \\
   1 & 1 & 0 \\   \end{array} } \right] \label{MRi}
\end{eqnarray}

The eigenvalues are

\begin{eqnarray}
M_R = \frac{f_V}{\sqrt{2}}  \left(\begin{array}{ccc} 2, & -1, & -1 \end{array}\right)
\end{eqnarray}

The corresponding eigenstates are


\begin{eqnarray}
N_R = \frac{1}{\sqrt{3}} (N^{(1)}) + N^{(2)} + N^{(3)} )_R \\ 
N^{\prime}_R = \frac{1}{\sqrt{2}} (N^{(1)}) - N^{(3)} )_R \\ 
N^{''}_R = \frac{1}{\sqrt{6}} (N^{(1)}) - 2 N^{(2)} + N^{(3)} )_R 
\end{eqnarray}

The group $SU(2) \times U(1)_Y \rightarrow U(1)_{em}$ is broken by a Higgs scalar doublet

\begin{eqnarray}
\phi = \left(\begin{array}{c} \phi^{+} \\ \phi^0 \end{array}\right) \rightarrow  \left(\begin{array}{c} 0 \\ \frac{H + v}{\sqrt{2}} \end{array}\right)
\end{eqnarray}

to give masses to the vector bosons $W^{\pm},~ Z_\mu$ and to quarks and charged leptons. The left-handed neutrinos in the standard model remain massless as there is no right-handed singlet neutrino to give the neutrino mass term.

In this model, there are three heavy right-handed neutrinos. Thus in this model the mass term for the neutrinos is generated by the Higgs doublet as follows

\begin{eqnarray}
\mathcal{L}_{mass} (\text{neutrino}) &=& \overline{L}_e h_1^{(2)} \left( \frac{\tau_1+i\tau_2}{2} \right)  \left(\begin{array}{c} 0 \\ \frac{H + v}{\sqrt{2}} \end{array}\right) N_R^{(1)}+\overline{L}_{\mu} h_2^{(3)} \left( \frac{\tau_1+i\tau_2}{2} \right)  \left(\begin{array}{c} 0 \\ \frac{H + v}{\sqrt{2}} \end{array}\right) N_R^{(3)}\notag\\
&~&~ + \overline{L}_{\tau} h_3^{(1)} \left( \frac{\tau_1+i\tau_2}{2} \right)  \left(\begin{array}{c} 0 \\ \frac{H + v}{\sqrt{2}} \end{array}\right) N_R^{(2)}
\end{eqnarray}

We note that Yukawa couplings $h^{(2)}_1, h^{(3)}_2, h^{(1)}_3$ of the Higgs doublet reflect the fact that $N^{(1)}_R$ connect the first generation with second,  $N^{(3)}_R$ second generation with third and  $N^{(2)}_R$ third generation with first. The mass matrix for the light neutrinos is given by 

\begin{eqnarray}
\mathcal{L}_{mass} (\text{neutrino}) &=& \frac{v}{\sqrt{2}} \left[ h_1^{(2)}\overline{\nu}_{eL}N_R^{(1)} + h_2^{(3)}\overline{\nu}_{\mu L}N_R^{(3)} + h_3^{(1)}\overline{\nu}_{\tau L}N_R^{(2)} + h.c. \right]\notag \\
&=& \frac{v}{\sqrt{2}} \left[ \frac{1}{\sqrt{3}} \left( h_1^{(2)}\overline{\nu}_{eL} + h_2^{(3)}\overline{\nu}_{\mu L} + h_3^{(1)}\overline{\nu}_{\tau L} \right) N_R + 
 \frac{1}{\sqrt{2}} \left( h_1^{(2)}\overline{\nu}_{eL} - h_2^{(3)}\overline{\nu}_{\mu L}  \right) N'_R  \right.\notag \\
&~&\qquad \left. +  \frac{1}{\sqrt{6}} \left( h_1^{(2)}\overline{\nu}_{eL} + h_2^{(3)}\overline{\nu}_{\mu L} - 2 h_3^{(1)}\overline{\nu}_{\tau L} \right) N''_R + h.c. \right] \label{Lnu}
\end{eqnarray}


From Eqs(\ref{MRi}) and (\ref{Lnu}), one gets the mass matrix for the light neutrinos \cite{1}.

\begin{eqnarray}
M_R = \frac{v^2}{2M_R}
  \left[ {\begin{array}{ccc}
   0 & h_1^{(2)}h_2^{(3)} & h_1^{(2)}h_3^{(1)} \\
   h_1^{(2)}h_2^{(3)} & 0 & h_2^{(3)}h_3^{(1)} \\
   h_1^{(2)}h_3^{(1)} & h_2^{(3)}h_3^{(1)} & 0 \\   \end{array} } \right]\label{MM}
\end{eqnarray}

where we have put $\frac{f_V}{\sqrt{2}} = M_R$


The three Yukawa couplings $h^{(2)}_1, h^{(3)}_2, h^{(1)}_3$ can be arranged in three different ways so as to give three dominant transitions as follows:
$$ \text{Case 1: } \nu_e \rightarrow \nu_\tau, \text{Case 2: } \nu_e \rightarrow \nu_\mu, \text{Case 3: }\nu_\mu \rightarrow \nu_\tau$$

Case $1, 2$ are relevant for the solar neutrino oscillations. The resonant amplification of solar neutrino oscillations in matter \cite{4,5} is an important aspect for the solar neutrino problem \cite{6}. The mass matrix (Eq. \ref{MM}) can accomodate this aspect of solar neutrino oscillations. Case $3$ is relevant for the atmospheric neutrino oscillations.

\subsection{Case $1$}
\begin{eqnarray}
 h^{(2)}_1 &=& h \sin\theta,~  h^{(1)}_3 = h \cos\theta  \notag \\
 \sigma &=& \frac{h}{h_2^{(3)}} \sin\theta \cos\theta,~m_0=\frac{v^2h h^{(3)}_2}{2M_R} \label{hs}
\end{eqnarray}

From Eq(\ref{MM}) and (\ref{hs}),

\begin{eqnarray}
   M = m_0
  \left[ {\begin{array}{ccc}
   0 &\sin \theta & \sigma \\
  \sin \theta & 0 &\cos \theta \\
   \sigma &\cos \theta & 0 \\   \end{array} } \right] \label{Mc1}
\end{eqnarray}

This mass matrix is similar to that first considered in \cite{7}. If $\sigma = 0$, the matrix (\ref{Mc1}) reduce to mass matrix discussed by several authors \cite{8}.

For the case $\sigma \ll \sin\theta$,

\begin{eqnarray}
   M^2 = \frac{1}{2} m^2_0
  \left[ {\begin{array}{ccc}
   1 & 0 & 0 \\
   0 & 1 & 0 \\
   0 & 0 & 1 \\   \end{array} } \right]
 + 
  \frac{1}{2} m^2_0
  \left[ {\begin{array}{ccc}
   -\cos 2\theta & 0 & \sin 2\theta \\
   0 & 1 & 0 \\
   \sin 2\theta & 0 & \cos 2\theta)  \end{array} } \right] \label{M21}
\end{eqnarray}

Thus, for Case $1$, dominant transition $\nu_e \rightarrow \nu_\tau$

\subsection{Case $2$}

\begin{eqnarray}
h^{(2)}_1 &=& h \sin\theta, ~ h^{(3)}_2 = h \cos\theta \notag \\
\sigma &=& \frac{h}{h_3^{(1)}}= \sin\theta \cos\theta, ~ m_0=\frac{v^2h h_3^{(1)}}{2M_R} \label{Mc2}
\end{eqnarray}
 
From Eq(\ref{MM}) and (\ref{Mc2}),

\begin{eqnarray}
   M = m_0
  \left[ {\begin{array}{ccc}

   0 & \sigma &\sin \theta  \\
   \sigma & 0 &\cos \theta \\
  \sin \theta &\cos \theta & 0 \end{array} } \right]
\end{eqnarray}
For $\sigma \ll\sin \theta$,

\begin{eqnarray}
   M^2 = \frac{1}{2} m^2_0
  \left[ {\begin{array}{ccc}
   1 & 0 & 0 \\
   0 & 1 & 0 \\
   0 & 0 & 1 \\   \end{array} } \right]
 + 
  \frac{1}{2} m^2_0
  \left[ {\begin{array}{ccc}
   -cos 2\theta &\sin 2\theta  & 0\\
   \sin 2\theta &\cos 2\theta  & 0\\
   0 & 0 & 1   \end{array} } \right] \label{M22}
\end{eqnarray}

For Case $2$, dominant transition $\nu_e \rightarrow \nu_\mu$


\subsection{Case $3$}

Here we consider two cases (a) and (b),

Case a) 

\begin{eqnarray}
h^{(1)}_3 &=& h\sin \theta, ~ h^{(3)}_2 = h\cos \theta \notag\\
\sigma &=& \frac{h}{h_1^{(2)}} \sin\theta \cos\theta, ~ m_0=\frac{v^2h h_1^{(2)}}{2M_R} \label{Mc3}
\end{eqnarray}

From Eq(\ref{MM}) and (\ref{Mc3}),

\begin{eqnarray}
   M = m_0
  \left[ {\begin{array}{ccc}
   0 &\cos \theta &\sin \theta  \\
  \cos \theta & 0 & \sigma \\
  \sin \theta & \sigma & 0 \\   \end{array} } \right]\label{M3}
\end{eqnarray}

Neglecting $\sigma,  \sigma \ll\sin \theta$,

\begin{eqnarray}
   M^2 = \frac{1}{2}
  \left[ {\begin{array}{ccc}
   1 & 0 & 0 \\
   0 & 1 & 0 \\
   0 & 0 & 1 \\   \end{array} } \right]
 + 
  \frac{1}{2} m^2_0
  \left[ {\begin{array}{ccc}
   1 & 0 & 0 \\
   0 &\cos 2\theta &\sin 2\theta) \\
   0 &\sin 2\theta & -cos 2\theta \\   \end{array} } \right] \label{M2a}
\end{eqnarray}

The dominant transition $\nu_\mu \rightarrow \nu_\tau$

Case b) 
\begin{eqnarray}
h^{(3)}_2 = h\sin \theta, ~ h^{(1)}_3 = h\cos \theta \label{Mc3b}
\end{eqnarray}

From Eq(\ref{M22}) and (\ref{Mc3b}), we get the mass matrix as given in Eq(\ref{M3}) with $cos \theta \leftrightarrow\sin \theta$


Neglecting $\sigma$, we get 

\begin{eqnarray}
   M^2 = \frac{1}{2} m^2_0
  \left[ {\begin{array}{ccc}
   1 & 0 & 0 \\
   0 & 1 & 0 \\
   0 & 0 & 1 \\   \end{array} } \right]
 + 
  \frac{1}{2} m^2_0
  \left[ {\begin{array}{ccc}
   1 & 0 & 0 \\
   0 & -cos 2\theta &\sin 2\theta) \\
   0 &\sin 2\theta &\cos 2\theta \\   \end{array} } \right] \label{M2b}
\end{eqnarray}

From Eqs. (\ref{M21}), (\ref{M22}) and (\ref{M2b}), it is clear that one can rewrite $M^2$ as a $2 \times 2$ matrix.

\begin{eqnarray}
   M^2 = \frac{1}{2} m^2_0\cos 2\theta
  \left[ {\begin{array}{cc}
   1 & 0 \\
   0 & 1 \\   \end{array} } \right]
 + 
  \frac{1}{2} m^2_0
  \left[ {\begin{array}{ccc}
   -2\cos 2\theta &\sin 2\theta) \\
   \sin 2\theta & 0 \\   \end{array} } \right] \label{M2}
\end{eqnarray}

which has the same form for Cases 1, 2 and 3b.

For 3a change, $-\cos 2\theta \leftarrow \cos 2\theta$

Now any $2\times2$ matrix 

\begin{eqnarray}
   M^2 =
  \left[ {\begin{array}{cc}
   a_1 & c \\
   c & a_2 \\   \end{array} } \right]
\end{eqnarray}

can be diagonalized by a unitary transformation
 
\begin{eqnarray}
   U =
  \left[ {\begin{array}{cc}
  \cos \theta_M &\sin \theta_M \\
   -sin \theta_M &\cos \theta_M \\   \end{array} } \right]
\end{eqnarray}



\begin{eqnarray}
   U^{-1} M^{2} U =
  \left[ {\begin{array}{cc}
  m_1^2 & 0 \\
  0 & m_2^2 \\   \end{array} } \right]
\end{eqnarray}
where

\begin{eqnarray}
\tan 2\theta_M &=& \frac{-2c}{a_1 - a_2} \label{thm} \\
m^{2}_{1,2} &=& M^{2}_{11,22} = \frac{1}{2} (a_1 + a_2) \pm \frac{1}{2} \frac{a_1 - a_2}{\cos 2\theta_M} \label{m12}
\end{eqnarray}

For the mass matrix $M^2$ in Eq.(\ref{M2})

\begin{eqnarray}
a_1 = -m^2_0\cos 2\theta , c = \frac{m^2_0\sin 2\theta}{2}
\end{eqnarray}

Hence from Eqs. (\ref{thm}, \ref{m12})

\begin{eqnarray}
\tan 2\theta_M = \tan 2\theta , \theta_M = \theta \\
m^2_1 + m^2_2 = -m_0\cos 2\theta \\
m^2_2 - m^2_1 = m^2_0 = \Delta m^2 \textgreater 0
\end{eqnarray}

so that Eq. (\ref{M2}) can be written in the form

\begin{eqnarray}
M^2 = \frac{1}{2} \Delta m^2\cos 2\theta \left[ {\begin{array}{cc} 1 & 0 \\ 0 & 1 \\  \end{array} } \right] + \left[ {\begin{array}{cc} - \Delta m^2\cos 2\theta & \frac{\Delta m^2\sin 2\theta}{2} \\ \\\frac{\Delta m^2\sin 2\theta}{2} & 0 \\  \end{array} } \right] \label{M2i}
\end{eqnarray}

But for the case 3a, $\theta_M = -\theta , \Delta m^2<0$.

However, for the solar neutrino oscillations, $\nu_e$ travelling in matter would acquire an extra effective mass $2\sqrt{2}G_F n_e E$, where E is the energy of neutrino and $n_e$ is the nuclear density of electrons and is given by \cite{6}

\begin{eqnarray}
n_e = (\frac{\rho}{m_N})Y
\end{eqnarray}

where Y denote the number of electrons per unit nucleon and is $\frac{1}{2}$ for ordinary matter.


Hence for the cases 1 and 2

\begin{eqnarray}
a_1 = - \Delta m^2 \cos 2\theta + 2\sqrt{2} G_F n_e E = -\Delta m^2 (cos 2\theta - A)
\end{eqnarray}

where 

\begin{eqnarray}
A = 2\sqrt{2} G_F n_e (\frac{E}{\Delta m^2})
\end{eqnarray}

Thus for the solar neutrino

\begin{eqnarray}
\tan 2\theta_M = \frac{sin 2\theta}{cos 2\theta -A} \\
\cos 2\theta_M = \frac{cos 2\theta - A}{[(cos 2\theta -A)^2+sin^2 2\theta]^{\frac{1}{2}}}\\
\sin 2\theta_M = \frac{sin 2\theta}{[(cos 2\theta -A)^2+sin^2 2\theta]^{\frac{1}{2}}}
\end{eqnarray}

Hence for Cases 1 and 2, the matrix $M^2$ given in Eq(\ref{M2i}) is modified

\begin{eqnarray}
M^2 = \frac{1}{2} \Delta m^2\cos 2\theta \left[ {\begin{array}{cc} 1 & 0 \\ 0 & 1 \\  \end{array} } \right] + \left[ {\begin{array}{cc} - \Delta m^2\cos 2 \theta + 2\sqrt{2} G_F n_e E& \frac{ \Delta m^2\sin 2\theta}{2} \\ \frac{\Delta m^2\sin 2\theta}{2} & 0 \\  \end{array} } \right]\label{M2f}
\end{eqnarray}



It is instructive to compare Eq(\ref{M2f}), with Hamiltonian for the solar neutrino oscillations \cite{2}

\begin{eqnarray}
H_M(E) = Const \left[ {\begin{array}{cc} 1 & 0 \\ 0 & 1 \\  \end{array} } \right] + \frac{1}{2E}\left[ {\begin{array}{cc} - \Delta m^2\cos 2\theta + 2\sqrt{2} G_F n_e E& \frac{\Delta m^2\sin 2\theta}{2} \\  \frac{\Delta m^2\sin 2\theta}{2} & 0 \\  \end{array} } \right]
\end{eqnarray}

The mass matrix $M^2$ has exactly the same form as that of $H_M(E)$ for solar neutrino oscillations \cite{2}. Thus it is relevant to compare predictions of our mass matrix with the experimental data from neutrino oscillations.

\section{NEUTRINO MASSES: COMPARISON WITH THE EXPERIMENTAL DATA}


Yukawa couplings $h^{(2)}_1, h^{(3)}_2, h^{(1)}_3$ for the three cases 1, 2 and 3 are analysed in detail.

Case 1: $\nu_e  \rightarrow \nu_\tau$

$$ h^{(2)}_1 = h\sin \theta,  h^{(1)}_3 = h\cos \theta$$ 

We assume 

\begin{eqnarray}
h^{(2)}_1 \frac{v}{\sqrt{2}} &=& \frac{1}{K} [2 m_e \sqrt{m_\mu m_\tau} ]^{\frac{1}{2}}\notag \\
h^{(1)}_3 \frac{v}{\sqrt{2}} &=& \frac{1}{K} [ m_\tau \sqrt{2 m_e m_\mu} ]^{\frac{1}{2}}\notag \\ 
h^{(3)}_2 \frac{v}{\sqrt{2}} &=& \frac{1}{K} \frac{m_\mu m_\tau}{\sqrt{m_e m_\mu}} 
\end{eqnarray}

The above choice gives

\begin{eqnarray}
\theta = 8.8^{\degree},~~ \sigma = 8.09 \times 10^{-4},~~\frac{1}{2}v^2h h_2^{(3)}=3.537 GeV^2/K^2
\end{eqnarray}
Thus
\begin{eqnarray}
\Delta m^2 > 0&,&\sin^2 \theta_{13} = 0.0234 \notag\\
\Delta m^2_{31} &=& 2.55\times 10^{-3} eV^2 \text{: } K^2M_R = 7 \times 10^{10} GeV 
\end{eqnarray}


Case $2$: $\nu_e \rightarrow \nu_\mu$

\begin{eqnarray}
 h^{(2)}_1 = h\sin \theta, ~ h^{(3)}_2 = h\cos \theta \notag
\end{eqnarray}

We assume 

\begin{eqnarray}
h^{(2)}_1 \frac{v}{\sqrt{2}} = \frac{1}{K} [\frac{m_e + m_\tau}{2} \sqrt{2m_e m_\mu} ]^{\frac{1}{2}} \notag\\ 
h^{(3)}_2 \frac{v}{\sqrt{2}} = \frac{1}{K} [\frac{m_\mu + m_e}{2} \sqrt{m_\mu m_\tau} ] \notag\\ 
h^{(1)}_3 \frac{v}{\sqrt{2}} = \frac{1}{K} \frac{m_\tau m_\mu}{\sqrt{m_\tau m_e}} 
\end{eqnarray}

This choice gives

\begin{eqnarray}
\theta &=& 32.2^{\degree},~~\sigma = 0.013 \\
\frac{v^2}{2}&h& h^{(1)}_3 = 1.126 GeV^2/K^2 \notag\\
\text{Thus, }&~&\Delta m^2 > 0,~~\sin^2 \theta_{12} = 0.285\notag\\
\Delta m^2_{21} &=& 7.50\times 10^{-3}eV^2:~K^2M_R = 1.3 \times 10^{11} GeV
\end{eqnarray}

Case $3$: $\nu_\mu \rightarrow \nu_\tau$ Atmospheric neutrino oscillations

a) \begin{eqnarray}
h^{(1)}_3 = h\sin \theta,  h^{(3)}_2 = h\cos \theta \notag
\end{eqnarray} 
we assume 
\begin{eqnarray}
h^{(1)}_3 \frac{v}{\sqrt{2}} = \frac{1}{K} (\frac{m_\tau + m_e}{2}) \notag\\ 
h^{(3)}_2 \frac{v}{\sqrt{2}} = \frac{1}{K} (\frac{m_\mu + m_\tau}{2}) \notag\\ 
h^{(2)}_1 \frac{v}{\sqrt{2}} = \frac{1}{K} \frac{m_\tau m_\mu}{\sqrt{2 m_e m_\mu}} 
\end{eqnarray}
This choice gives
\begin{eqnarray}
\theta &=& 43.35^{\degree},~~\sigma = 0.035,~~\frac{v^2}{2} h h^{(2)}_1=23.36 GeV^2/K^2  \\
\text{Thus }&~&\Delta m^2<0,~~\sin^2\theta_{23}=0.471\notag \\
\mid\Delta m^2\mid &=& \Delta m^2_{23} =2.47\times 10^{-3}eV^2 :~~K^2M_R = 4.7 \times 10^{11} GeV
\end{eqnarray}

b)\begin{eqnarray}
h^{(3)}_2 = h\sin \theta,  h^{(1)}_3 = h\cos \theta \notag
\end{eqnarray}

we assume 

\begin{eqnarray}
h^{(3)}_2 \frac{v}{\sqrt{2}} &=& \frac{1}{K} (\frac{1}{2} (m_\mu + m_e) \sqrt{m_\tau  m_\mu})^{\frac{1}{2}} \notag\\ 
h^{(1)}_3 \frac{v}{\sqrt{2}} &=& \frac{1}{K} (\frac{1}{2} (m_\tau + m_\mu) \sqrt{m_\tau  m_e})^{\frac{1}{2}} \notag\\ 
h^{(2)}_1 \frac{v}{\sqrt{2}} &=& \frac{1}{K} \frac{m_\mu m_\tau}{\sqrt{2 m_e m_\mu}} 
\end{eqnarray}

This choice gives
\begin{eqnarray}
\theta = 42.07^{\degree}&,& \sigma = 6.22 \times 10^{-3},~~\frac{v^2}{2} h h^{(2)}_1=4.09 GeV^2/K^2 \notag\\ 
\text{Thus }&~&\Delta m^2 > 0,\sin^2 \theta_{23} = 0.450 \notag \\
\Delta m^2 &=& \Delta m^2_{32} = 2.49 \times 10^{-3} eV^{2} : K^2 M_R = 8.2 \times 10^{10} GeV
\end{eqnarray}

%

\section{Summary}
To summarize: The Yukawa couplings $h^{(2)}_1, h^{(3)}_2, h^{(1)}_3$ are arranged and expressed in terms of masses $m_e$, $m_\mu$ and $m_\tau$ to obtain results of interest for three cases.

Case 1: ($\nu_e \rightarrow \nu_\tau$), Case 2: ($\nu_e \rightarrow \nu_\mu$), Case 3: ($\nu_\mu \rightarrow \nu_\tau$). For all three casse analysed $\sigma$ is negligible; as consequence of this, the neutrino mass matrix can be expressed as a $2\times 2$ matrix.

For Cases 1 and 2, our analysis gives 
\begin{eqnarray}
\Delta m^2>0,~\sin^2 \theta_{13} = 0.0234, ~ \Delta m_{31}^2\approx 2.55\times 10^{-3} eV^2,\notag \\ 
\sin^2 \theta_{12} = 0.285, ~  \Delta m_{21}^2\approx 7.50\times 10^{-5} eV^2.\notag 
\end{eqnarray}

For Case 3: 
\begin{eqnarray}
a)~~\Delta m^2 < 0&,&~\sin^2 \theta_{23} \approx 0.471,~
|\Delta m^2|= \Delta m_{23}^2\approx 2.47\times 10^{-3} eV^2\notag \\
b)~~\Delta m^2 > 0&,&~\sin^2 \theta_{23} \approx 0.450, ~
\Delta m_{32}^2\approx 2.49\times 10^{-3} eV^2\notag
\end{eqnarray}
to be compared with experimental values from neutrino oscillation experiments ($\nu_e \rightarrow \nu_\tau$), ($\nu_e \rightarrow \nu_\mu$), ($\nu_\mu \rightarrow \nu_\tau$)\cite{3}.

Best fit ($1\sigma$)
\begin{eqnarray}
\Delta m^2 > 0,&~&\sin^2 \theta_{13} \approx 0.0214,\notag\\
			&~&\sin^2 \theta_{12} \approx 0.297,~ \Delta m_{21}^2\approx 7.37\times 10^{-5} eV^2 \notag\\
\Delta m^2 < 0,&~&\sin^2 \theta_{23} \approx 0.569,\\ &~&\qquad\qquad \Delta m_{32}^2=|\Delta m_{23}^2|\approx 2.50\times 10^{-3} eV^2 \notag\\
\Delta m^2 > 0,&~&\sin^2 \theta_{23} \approx 0.437~.\notag
\end{eqnarray}
Except for $\Delta m^2<0$, $\sin^2 \theta_{23}$, our values agree with the above values within few percent and are well within $3\sigma$ fit.

Final remark: In our analysis we have used normal hierarch $m_1<m_2<m_3$; thus
\begin{eqnarray}
\Delta m^2&=&|\Delta m_{32}^2|\approx 2.50\times 10^{-3} eV^2 \notag\\
\Delta m_{31}^2 &\approx& \Delta m_{32}^2 \notag\\
\frac{\Delta m_{21}^2}{\Delta m^2}&\approx& 0.030\notag\\
\text{Experimental value}&=&0.029\notag
\end{eqnarray}
To get $\Delta m^2$ within the experimental range, one would require $K^2M_R\approx10^{11}$GeV.

\end{document}